\documentclass[twocolumn,showpacs,showkeys,reprint,amsmath,amssymb,aps]{revtex4-1}

\usepackage{graphicx}
\usepackage{dcolumn}
\usepackage{bm}
\usepackage{tabularx}
\usepackage{amsmath}

\begin{document}

\title{Absence of magnetic thermal conductivity in the quantum spin liquid candidate EtMe$_3$Sb[Pd(dmit)$_2$]$_2$ -- revisited}

\author{J. M. Ni,$^1$ B. L. Pan,$^1$ Y. Y. Huang,$^1$ J. Y. Zeng,$^1$ Y. J. Yu,$^1$ E. J. Cheng,$^1$ L. S. Wang,$^1$ R. Kato,$^2$ and S. Y. Li$^{1,3,*}$}

\affiliation
{$^1$State Key Laboratory of Surface Physics, Department of Physics, and Laboratory of Advanced Materials, Fudan University, Shanghai 200438, China\\
 $^2$RIKEN, Condensed Molecular Materials Laboratory, Wako 351-0198, Japan\\
 $^3$Collaborative Innovation Center of Advanced Microstructures, Nanjing 210093, China
}

\date{\today}

\begin{abstract}
We present the ultralow-temperature specific heat and thermal conductivity measurements on single crystals of triangular-lattice organic compound EtMe$_3$Sb[Pd(dmit)$_2$]$_2$, which has long been considered as a gapless quantum spin liquid candidate. In specific heat measurements, a finite linear term is observed, consistent with the previous work [S. Yamashita $et$ $al.$, Nat. Commun. {\bf 2}, 275 (2011)]. However, we do not observe a finite residual linear term in the thermal conductivity measurements, and the thermal conductivity does not change in a magnetic field of 6 Tesla. These results are in sharp contrast to previous thermal conductivity measurements on EtMe$_3$Sb[Pd(dmit)$_2$]$_2$ [M. Yamashita $et$ $al.$ Science {\bf 328}, 1246 (2010)], in which a huge residual linear term was observed and attributed to highly mobile gapless excitations, likely the spinons of a quantum spin liquid. In this context, the true ground state of EtMe$_3$Sb[Pd(dmit)$_2$]$_2$ has to be reconsidered.
\end{abstract}

\maketitle

Quantum spin liquid (QSL) states have been one of the central issues in condensed matter physics since the seminal proposal by Anderson \cite{anderson73,anderson87}. Due to the strong geometrical frustration and quantum fluctuations, the spins do not order even down to zero temperature and remain highly entangled, and fractionalized excitations called spinons are the most pronounced characteristic in QSL states \cite{balents10,savary17,zhourmp}. The detection and study of these excitations are of great importance to give information about the nature of QSL states. As the prototype of a QSL in Anderson's resonating valence bond (RVB) model \cite{anderson73}, the spin 1/2 triangular-lattice Heisenberg antiferromagnet is a platform for searching the QSL candidates. Among those triangular-lattice QSL candidates, the two organic compounds $\kappa$-(BEDT-TTF)$_2$Cu$_2$(CN)$_3$ and EtMe$_3$Sb[Pd(dmit)$_2$]$_2$ seem to be promising \cite{ETprl03,ETprb06,ETC,ETk,dmitJPCM,dmitprb08,dmitk,dmitNMR1,dmitNMR2,dmitC,dmittorque}, while the inorganic compound YbMgGaO$_4$ is under hot debate  \cite{YMGOsr,YMGOneutron1,YMGOneutron2,YMGOk,YZGO}.

EtMe$_3$Sb[Pd(dmit)$_2$]$_2$ (Me = CH$_3$, Et = C$_2$H$_5$, dmit = 1,3-dithiole-2-thione-4,5-dithiolate) is a compound of the series of layered organic salts $\beta$'-X[Pd(dmit)$_2$]$_2$, where X is a nonmagnetic cation, and Pd(dmit)$_2$ is highly dimerized forming the spin 1/2 anion [Pd(dmit)$_2$]$_2$$^-$ \cite{katoreview04,katoreview11}. Each Pd(dmit)$_2$ layer is parallel to the $ab$ plane and is separated by the EtMe$_3$Sb$^+$ cation layer, as seen in Fig. 1(a) (cation layers are not shown for clarity). The dimerized Pd(dmit)$_2$ are arranged to form a quasi-triangular lattice in the layer, causing a strong geometrical frustration of spins on the dimers, illustrated in Fig. 1(b). No signature of long range magnetic order was observed down to about 20 mK by nuclear magnetic resonance (NMR) measurements \cite{dmitNMR1,dmitNMR2}, in spite of the large effective antiferromagnetic exchange interactions of the order of 250 K \cite{dmitprb08}. However, inhomogeneity gradually develops on cooling \cite{dmitNMR1,dmitNMR2}. As for the study of excitations in this putative QSL candidate, the spin-lattice relaxation $1/T_1$ was found to follow the $T^2$ temperature dependence below 1 K, indicating a nodal spin gap \cite{dmitNMR1,dmitNMR2}. In contrast, a finite linear term $\gamma$ of 19.9 mJ K$^{-2}$ mol$^{-1}$ was observed in the specific heat measurements, implying the existence of gapless fermionic excitations \cite{dmitC}. The gapless nature was further confirmed by the torque magnetometry, which revealed residual paramagnetic susceptibility comparable to that in metal \cite{dmittorque}.

Ultralow-temperature thermal conductivity measurement is a very useful technique to study the low-lying excitations, even for charge-neutral quasiparticles. The previous thermal conductivity work of EtMe$_3$Sb[Pd(dmit)$_2$]$_2$ single crystals by M. Yamashita $et$ $al.$ reported the observation of a huge residual linear term $\kappa_0/T$ of 2 mW K$^{-2}$ cm$^{-1}$, which is attributed to highly mobile gapless fermionic excitations with the mean free path as long as 1000 inter-spin distances \cite{dmitk}. This is a strong evidence for the existence of a spinon Fermi surface in such a QSL candidate, stimulating a number of theoretical studies \cite{savary17,zhourmp}. The enhancement of $\kappa$ above 2 T was also observed \cite{dmitk}. However, for other two triangular-lattice QSL candidates $\kappa$-(BEDT-TTF)$_2$Cu$_2$(CN)$_3$ and YbMgGaO$_4$, no residual linear terms were found in thermal conductivity measurements \cite{ETk,YMGOk}, although they also display gapless nature in specific heat measurements (i.e. a linear term $\gamma$ of 15 mJ K$^{-2}$mol$^{-1}$ in $\kappa$-(BEDT-TTF)$_2$Cu$_2$(CN)$_3$ and a power-law temperature dependence in YbMgGaO$_4$, respectively) \cite{ETC,YMGOsr,YMGOk}. One may raise the question why EtMe$_3$Sb[Pd(dmit)$_2$]$_2$ is so different from other triangular-lattice QSL candidates. Therefore, as the vital experimental foundation of many theoretical works, it is desired to revisit these thermodynamic and transport properties of EtMe$_3$Sb[Pd(dmit)$_2$]$_2$.

In this Letter, we report the ultralow-temperature specific heat and thermal conductivity measurements on high-quality EtMe$_3$Sb[Pd(dmit)$_2$]$_2$ single crystals. A linear term $\gamma$ of 14.9 mJ K$^{-2}$ cm$^{-1}$ is found in the specific heat, which is consistent with the previous work \cite{dmitC}. However, it is unsuccessful to reproduce previous thermal conductivity results reported in Ref. \cite{dmitk}. Negligible residual linear term $\kappa_0/T$ is observed, implying the absence of mobile gapless fermionic excitations. This raises the question about the true ground state of this triangular-lattice QSL candidate.

\begin{figure}
\includegraphics[clip,width=8cm]{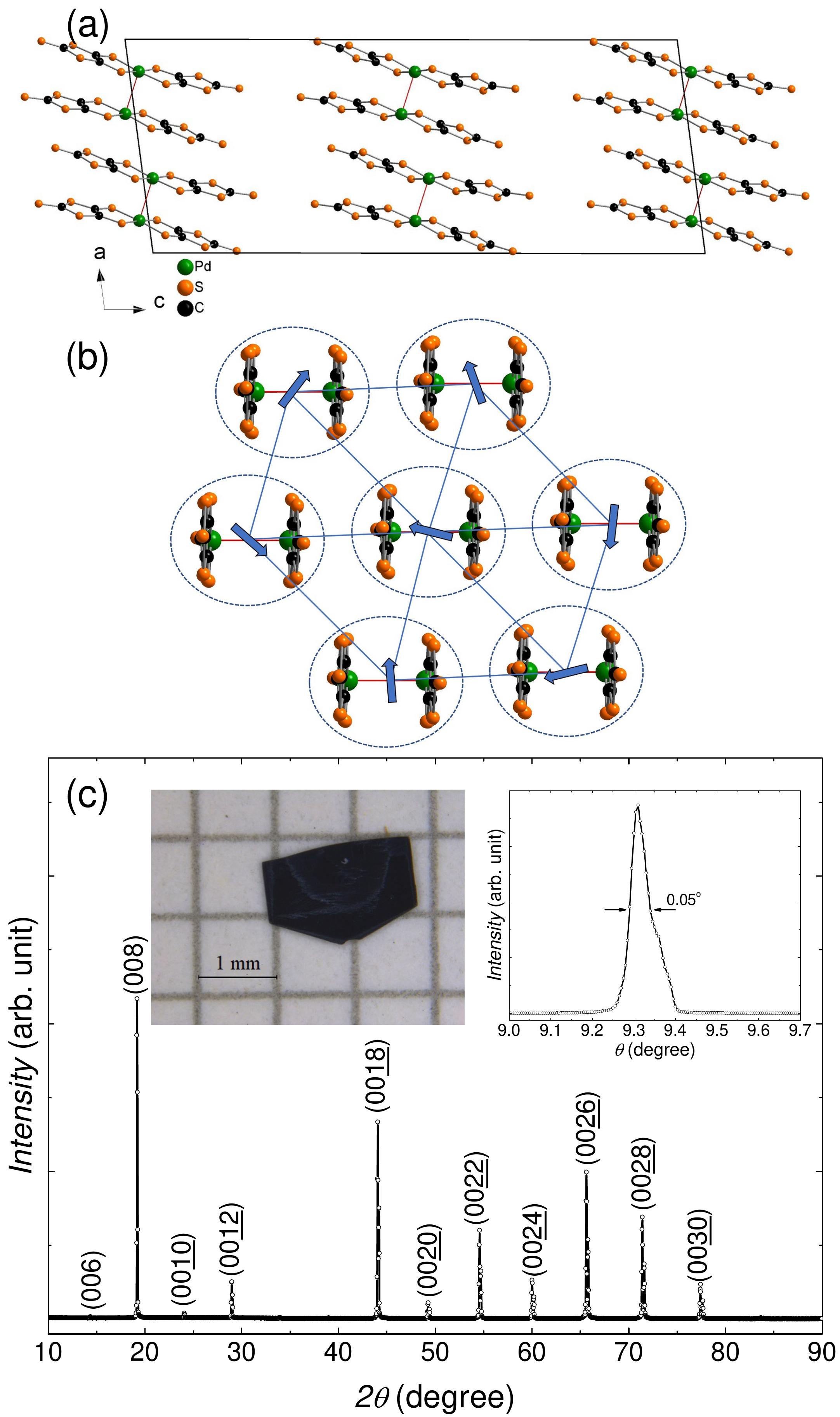}
\caption{(a) Crystal structure of EtMe$_3$Sb[Pd(dmit)$_2$]$_2$ with only the Pd(dmit)$_2$ layers shown. (b) The triangular-lattice structure of the spin 1/2 in the Pd(dmit)$_2$ layer. The dimerized Pd(dmit)$_2$ are surrounded by dashed ovals. Each 1/2-spin, denoted as a blue arrows, is localized on each dimer. (c) Room-temperature XRD pattern from the natural surface of the EtMe$_3$Sb[Pd(dmit)$_2$]$_2$ single crystal. Left inset: the photo of a typical EtMe$_3$Sb[Pd(dmit)$_2$]$_2$ single crystal. Right inset: x-ray rocking scan curve of the (008) Bragg peak.}
\end{figure}

\begin{figure}
\includegraphics[clip,width=8cm]{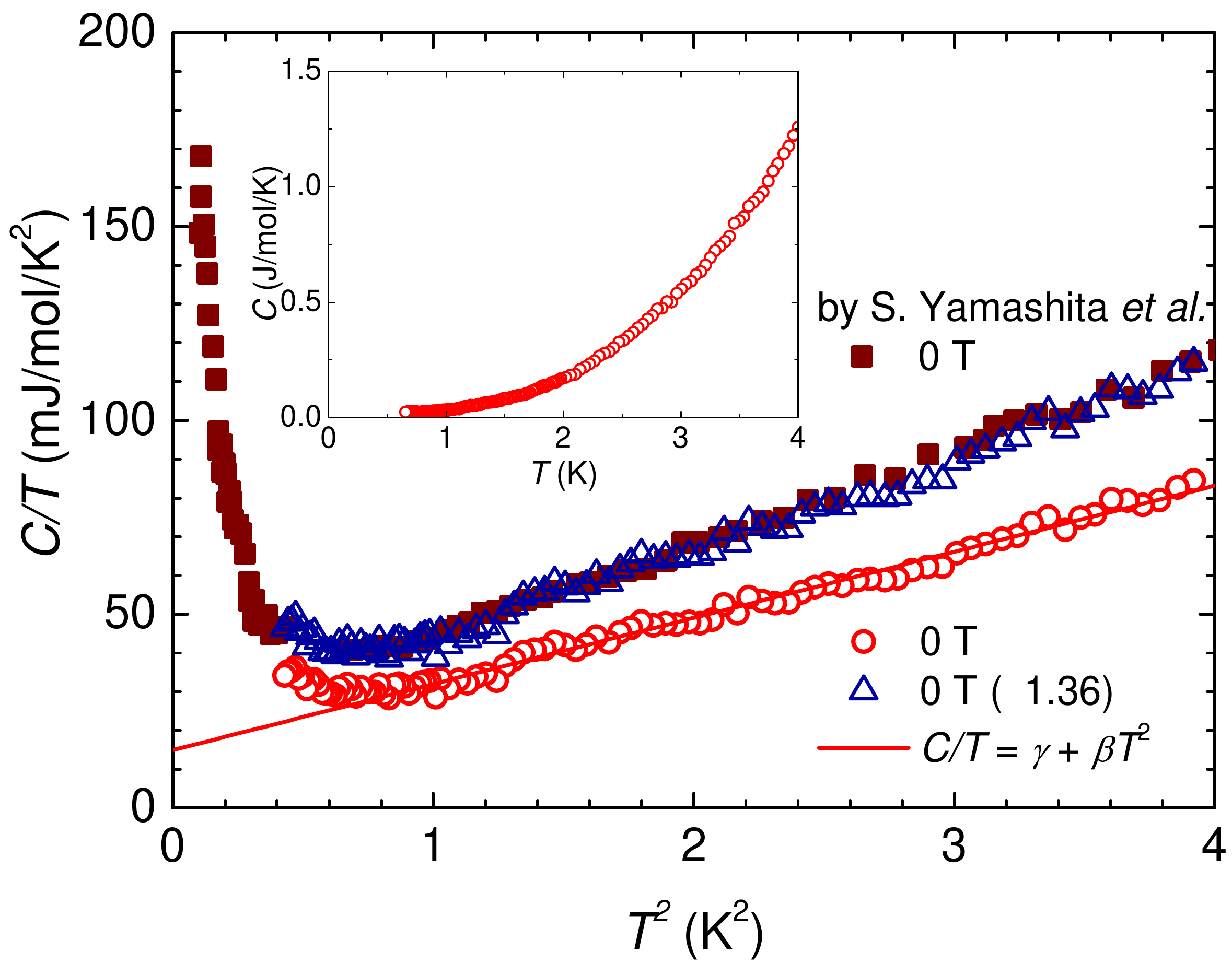}
\caption{Temperature dependence of the specific heat of EtMe$_3$Sb[Pd(dmit)$_2$]$_2$ single crystal at zero field. The inset shows $C$ vs $T$ from 0.65 to 4 K. The main panel plots $C/T$ vs $T^2$ below 2 K. The solid line is the fit to the specific heat data between 0.9 K and 2 K to $C/T$ = $\gamma$ + $\beta$$T^2$. A linear term $\gamma$ = 14.9 $\pm$ 0.5 mJ K$^{-2}$ mol$^{-1}$ is obtained. For comparison, the specific heat result in Ref. \cite{dmitC} is also plotted. Scaling by a factor of 1.36 can well reproduce the prior result, indicating that the difference is likely due to the uncertainty in measuring the mass of the samples.}
\end{figure}

Single crystals of EtMe$_3$Sb[Pd(dmit)$_2$]$_2$ were grown by air oxidation of (EtMe$_3$Sb)$_2$[Pd(dmit)$_2$] (60 mg) in acetone (100 ml) containing acetic acid (7-10 ml) at low temperatures in the range of -11 to 5 $^\circ$C, provided by the same group at RIKEN as in Ref. \cite{dmitk}. The photo of a typical sample is shown in the left inset of Fig. 1(c). The x-ray diffraction (XRD) measurement was performed on a EtMe$_3$Sb[Pd(dmit)$_2$]$_2$ sample by using an x-ray diffractometer (D8 Advance, Bruker), and determined the natural surface to be (00$l$) plane, as seen in Fig. 1(c). The quality of EtMe$_3$Sb[Pd(dmit)$_2$]$_2$  single crystals was checked by the x-ray rocking scan, shown in the right inset of Fig. 1(c). The full width at half-maximum (FWHM) is only 0.05$^\circ$, indicating the high quality of the samples. The specific heat of a sample with 0.4 $\pm$ 0.1 mg was measured by the relaxation method in a physical property measurement system (PPMS, Quantum Design) equipped with a $^3$He cryostat. Samples for thermal conductivity measurements have the dimensions of 1.34 $\times$ 0.60 $\times$ 0.05 mm$^3$ for sample $\#$1, 2.00 $\times$ 1.17 $\times$ 0.05 mm$^3$ for sample $\#$2, and 1.34 $\times$ 0.88 $\times$ 0.02 mm$^3$ for sample $\#$3, respectively. Note that the samples used for thermal conductivity measurements come from three different batches. All samples were prepared in almost same conditions as samples in Ref. \cite{dmitk}. As for sample $\#$3, especially, the conditions were the same including the reagents. For EtMe$_3$Sb[Pd(dmit)$_2$]$_2$ single crystals, no change in properties was observed with time. Contacts of sample $\#$1 were made by gold paste (PELCO Conductive Gold Paste, Product No 16022) with the thinner of Diethyleneglycol monoethyl ether. Contacts of samples $\#$2 and $\#$3 were made by carbon paste (Dotite paint XC-12 from JEOL) with the thinner of Dimethyl Succinate. The thermal conductivity was measured in a dilution refrigerator, using a standard four-wire steady-state method with two RuO$_2$ chip thermometers, calibrated $in$ $situ$ against a reference RuO$_2$ thermometer. Samples were cooled slowly from room temperature to the lowest temperature for 2 days in order to avoid cracks. Magnetic fields were applied perpendicular to $ab$ plane.

The temperature dependence of the specific heat of EtMe$_3$Sb[Pd(dmit)$_2$]$_2$ single crystal from 0.65 to 4 K at zero field is shown in the inset of Fig. 2. The main panel of Fig. 2 plots the $C/T$ vs $T^2$ below 2 K. The upturn at low temperatures is attributed to the Schottky anomaly which comes from the rotational tunneling of methyl groups in the cation layer \cite{dmitC}. $C/T$ can be well fitted by the formula $C/T$ = $\gamma$ + $\beta$$T^2$ between 0.9 K and 2 K, giving $\gamma$ = 14.9 $\pm$ 0.5 mJ K$^{-2}$ mol$^{-1}$ and $\beta$ = 17.1 $\pm$ 0.2 mJ K$^{-4}$ mol$^{-1}$. This behavior is similar to that in Ref. \cite{dmitC}, which is also plotted Fig. 2. Actually, if we scale our data by a factor of 1.36, they are on top of the curve in Ref. \cite{dmitC}. Thus the previous specific heat result is well reproduced by us, and the slight difference is likely due to the uncertainty in determining the mass of samples. The finite linear term $\gamma T$ in the specific heat, which is unexpected for an insulator, was considered as the evidence of gapless excitations with fermionic statistics \cite{dmitC}.

\begin{figure}
\includegraphics[clip,width=8cm]{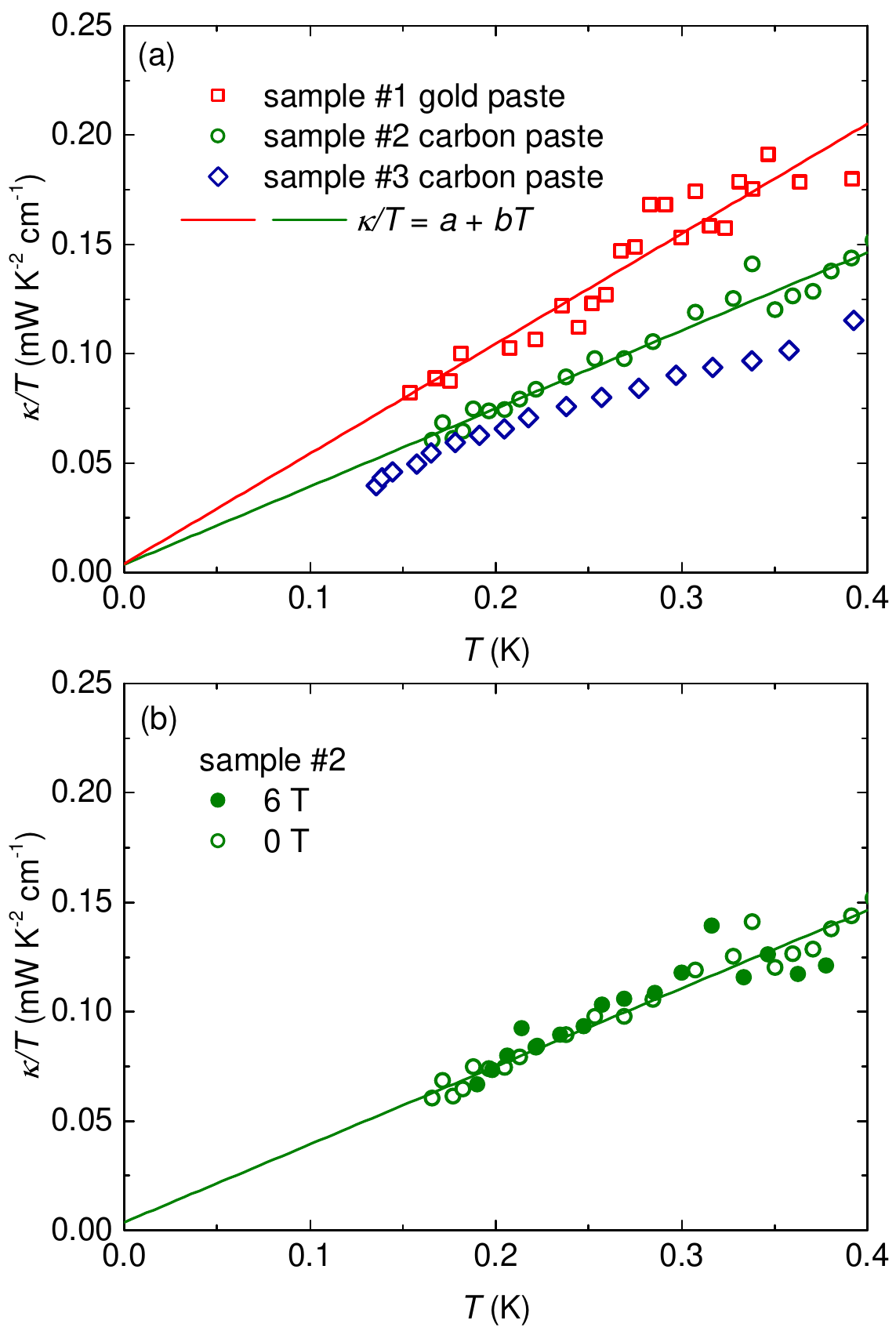}
\caption{(a) The zero-field in-plane thermal conductivity of three EtMe$_3$Sb[Pd(dmit)$_2$]$_2$ single crystals, using contact of gold paste and carbon paste, respectively. (b) The in-plane thermal conductivity of sample $\#$2 at $\mu_0H$ = 0 T and 6 T. The straight lines are the fits to the zero-field thermal conductivity data to $\kappa/T$ = $a$ + $bT$.}
\end{figure}

Since specific heat result may be contaminated by nuclear contributions \cite{ETC,dmitC}, the thermal transport measurement, a tool insensitive to localized excitations, would be more advantageous to identify the low-energy excitations in a QSL candidate. Figure 3(a) presents the in-plane zero-field thermal conductivity of EtMe$_3$Sb[Pd(dmit)$_2$]$_2$ single crystals with two kinds of contacts, respectively made with gold paste and carbon paste. The thermal conductivity of an insulating QSL candidate at very low temperatures usually can be fitted by $\kappa/T$ = $a$ + $bT^{\alpha-1}$, in which the two terms $aT$ and $bT^\alpha$ represent the contributions from itinerant gapless fermionic magnetic excitations (if they do exist) and phonons, respectively. For phonons, the power $\alpha$ is typically between 2 and 3, due to the specular reflections at the sample surfaces \cite{fit1,fit2}. As can be seen in Fig. 3(a), contrary to Ref. \cite{dmitk} where $\kappa/T$ shows $T^2$ temperature dependence, a linear fitting $\kappa/T$ = $a$ + $bT$ is more suitable for our samples $\#$1 and $\#$2. The fitting gives the value of $\kappa_0/T$ of 0.004 $\pm$ 0.009 mW K$^{-2}$ cm$^{-1}$ and 0.004 $\pm$ 0.005 mW K$^{-2}$ cm$^{-1}$ for sample $\#$1 and sample $\#$2, respectively. Considering the experimental error bar of 5 $\mu$W K$^{-2}$ cm$^{-1}$, the $\kappa_0/T$ of both samples at $\mu_0H$ = 0 T are virtually zero. To further confirm our results, we also measure sample $\#$3 prepared from the same conditions including the reagents as those in Ref. \cite{dmitk}. The behavior is very similar to sample $\#1$ and $\#2$. Due to the slightly sublinear temperature dependence, it extrapolates to a small negative value of $\kappa_0/T$. Therefore previous huge $\kappa_0/T$ of EtMe$_3$Sb[Pd(dmit)$_2$]$_2$ in Ref. \cite{dmitk} can not be reproduced in any of our samples.

The in-plane thermal conductivities of sample $\#$2 at $\mu_0H$ = 0 T and 6 T are plotted in Fig. 3(b). They basically overlap with each other. In other words, the magnetic field barely has any effect on the thermal conductivity of EtMe$_3$Sb[Pd(dmit)$_2$]$_2$ up to 6 T. This magnetic field dependence, again, is in stark contrast to the previous measurements that report a gradual increase of thermal conductivity above approximately 2 T below 1 K \cite{dmitk}. For example, the observed magneto-thermal conductivity $(\kappa(H)-\kappa(0$T$))/\kappa(0$T$)$ is larger than 20\% at 6 T at 0.23 K, which was suggested as a consequence of additional excitations with a spin gap closing at 2 T \cite{dmitk}. Due to the coexistence of gapless and gapped excitations, a type-II spin liquid was proposed for EtMe$_3$Sb[Pd(dmit)$_2$]$_2$ \cite{RPP11}. Considering the negligible field effect presented here, one may reexamine the existence of this gapped excitations and the proposal of type-II spin liquid.

\begin{figure}
\includegraphics[clip,width=8cm]{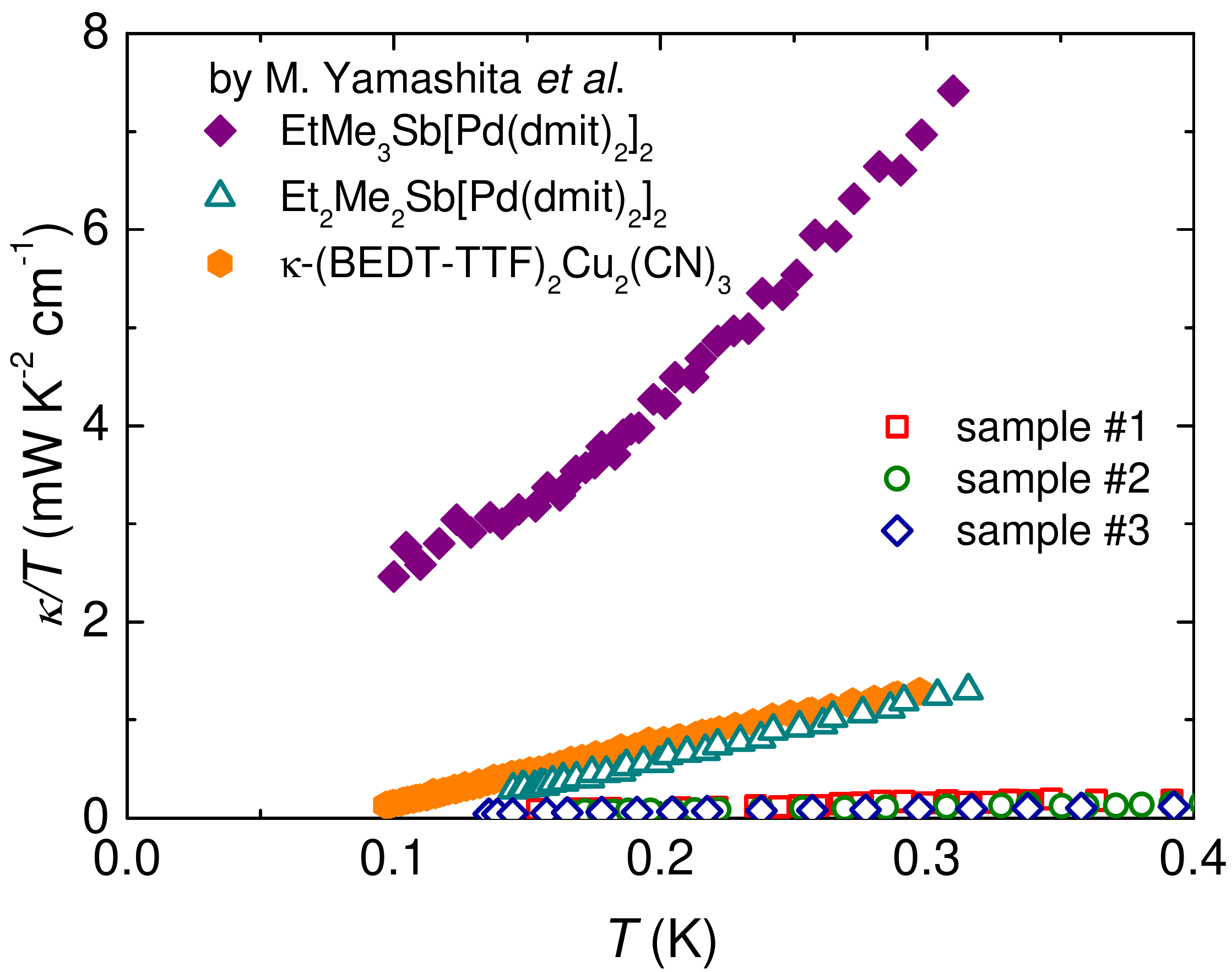}
\caption{Comparison of our thermal conductivity data with the data of EtMe$_3$Sb[Pd(dmit)$_2$]$_2$ and Et$_2$Me$_2$Sb[Pd(dmit)$_2$]$_2$ in Ref. \cite{dmitk} and $\kappa$-(BEDT-TTF)$_2$Cu$_2$(CN)$_3$ in Ref. \cite{ETk}. The absolute value of our data are much smaller than that in Ref. \cite{dmitk}, even 10 times smaller than the nonmagnetic reference compound Et$_2$Me$_2$Sb[Pd(dmit)$_2$]$_2$, indicating phonons being strongly scattered in our samples.}
\end{figure}

Figure 4 displays the comparison of our thermal conductivity data with the data of EtMe$_3$Sb[Pd(dmit)$_2$]$_2$ and Et$_2$Me$_2$Sb[Pd(dmit)$_2$]$_2$ in Ref. \cite{dmitk} and $\kappa$-(BEDT-TTF)$_2$Cu$_2$(CN)$_3$ in Ref. \cite{ETk}. The absolute value of our data are much smaller than that in Ref. \cite{dmitk}, even 10 times smaller than the nonmagnetic reference compound Et$_2$Me$_2$Sb[Pd(dmit)$_2$]$_2$ \cite{dmitk}. The lack of $\kappa_0/T$, negligible field effect in $\mu_0H$ = 6 T, and very small absolute value, all these results demonstrate that the thermal conductivity is entirely contributed by phonons in our samples, and the phonons are strongly scattered. This is supported by the estimation of the mean free path $l_p$ of phonons by the kinetic formula $\kappa$ = $\frac{1}{3}$${C_p}$$v_p$$l_p$, where $C_p$ = $(2\pi^2k_B/5)$$(k_BT/\hbar v_p)^3$ = $\beta T^3$ is the phonon specific heat and $v_p$ is the velocity of phonons. With $\beta$ = 17.1 mJ K$^{-4}$ mol$^{-1}$ obtained by our specific heat measurement, $v_p$ is estimated as 1.55 $\times$ 10$^3$ m/s, giving $l_p$ of sample $\#$3 only about 5.88 $\mu$m at 0.3 K. Usually when the sample enters boundary scattering limit at sub-Kelvin temperature, $l_p$ should only be limited by the physical dimensions of the sample, resulting in a temperature independent $l_p$ = 2$\sqrt{A/\pi}$, where $A$ is the cross-sectional area of the sample \cite{fit2}. For sample $\#$3, the boundary limited $l_p$ is 150 $\mu$m. This is 25 times larger than $l_p$ estimated from our thermal conductivity data, and consistent with the fact that the thermal conductivity of magnetic EtMe$_3$Sb[Pd(dmit)$_2$]$_2$ is 10 times lower than the nonmagnetic Et$_2$Me$_2$Sb[Pd(dmit)$_2$]$_2$. It is likely that the phonons are strongly scattered by those frustrated spins in EtMe$_3$Sb[Pd(dmit)$_2$]$_2$. Note that in another triangular-lattice system, the thermal conductivity of magnetic YbMgGaO$_4$ is about half of the nonmagnetic LuMgGaO$_4$, again showing the scattering of phonons by the spins \cite{YMGOk}. There is a difference in the field dependence of the phonon thermal conductivity. In YbMgGaO$_4$, $\kappa$ is enhanced by about 25\% in $\mu_0H >$ 5 T, due to the suppression of phonon-spin scattering \cite{YMGOk}. In EtMe$_3$Sb[Pd(dmit)$_2$]$_2$, the field effect on thermal conductivity is negligible in $\mu_0H$ = 6 T, which is likely due to its much higher antiferromagnetic exchange interactions than YbMgGaO$_4$.

Now we would like to discuss the implications of the negligible $\kappa_0/T$ in EtMe$_3$Sb[Pd(dmit)$_2$]$_2$ according to our new results. A gapless QSL may have fermionic spinons, which can form a Fermi surface like a metal, and provide a large density of states at low energies. Therefore, a sizable $\kappa_0/T$ is expected \cite{ktheory1,ktheory2,ktheory4,ktheory3}. In Ref. \cite{dmitk}, from the observed huge $\kappa_0/T$ = 2 mW K$^{-2}$ cm$^{-1}$, the authors estimated the mean free path $l_s$ of gapless fermionic excitations as long as 1000 inter-spin distances. However, according to the fitting result $\kappa_0/T$ = 0.004 mW K$^{-2}$ cm$^{-1}$ of our samples $\#$1 and $\#$2, an upper bound of $l_s$ around 10 \AA\ is estimated, comparable to the inter-spin distance. In this context, the negligible $\kappa_0/T$ observed in our experiments is inconsistent with the existence of highly mobile gapless fermionic excitations and spinon Fermi surface in EtMe$_3$Sb[Pd(dmit)$_2$]$_2$. The reason for such a huge discrepancy between our results and those in Ref. \cite{dmitk} is not clear for us, but it is clearly not due to sample dependence since we have achieved similar results in three samples from different batches, especially including one prepared from the same conditions including the reagents as those in Ref. \cite{dmitk}. Below we only discuss the feasibility of possible scenarios in light of our experimental data on low-lying excitations.

One possible scenario is that EtMe$_3$Sb[Pd(dmit)$_2$]$_2$ has a fully gapped ground state. This scenario was initially used to explain the absence of $\kappa_0/T$ in $\kappa$-(BEDT-TTF)$_2$Cu$_2$(CN)$_3$ \cite{ETk}. By fitting the sublinear temperature-dependent $\kappa/T$ at low temperature, a gap of about 0.46 K was obtained \cite{ETk}. However, such a gapped scenario is inconsistent with the finite linear term in the specific heat. Therefore, the authors in Ref. \cite{ETk} pointed out that the heat capacity measurements incorrectly suggest the presence of gapless excitation, possibly owing to the large Schottky contribution at low temperatures. Now that the similar phenomena appear in EtMe$_3$Sb[Pd(dmit)$_2$]$_2$ according to our thermodynamic and heat transport results, one may take both two compounds into account for this scenario.

Another possible scenario is that gapless excitations do exist in EtMe$_3$Sb[Pd(dmit)$_2$]$_2$ and $\kappa$-(BEDT-TTF)$_2$Cu$_2$(CN)$_3$, indicated by specific heat measurements, but they are localized. Organic molecular compounds are generally thought as extremely clean with a very low level of crystallographic defects. However, the exponent in the stretched-exponential fitting of the relaxation curve in NMR measurements of $\kappa$-(BEDT-TTF)$_2$Cu$_2$(CN)$_3$ deviates from unity substantially below 6 K, indicating some inhomogeneity therein \cite{ETprb06}. This scenario was considered as an alternative explanation for the negligible $\kappa_0/T$ in $\kappa$-(BEDT-TTF)$_2$Cu$_2$(CN)$_3$ \cite{kreview}. For EtMe$_3$Sb[Pd(dmit)$_2$]$_2$, the inhomogeneity was observed below 10 K in NMR measurements \cite{dmitNMR1,dmitNMR2}. Although the stretching exponents which characterize the degree of inhomogeneity start to recover towards the homogeneous value below 1 K, they are only about 0.6 which still deviates from the value of a homogeneous system \cite{dmitNMR1,dmitNMR2}. Thus, the inhomogeneity still cannot be ruled out in sub-Kelvin temperature region \cite{zhourmp}. Therefore, our thermal conductivity results raise the question about to what extent does the inhomogeneity play a role in the heat transport of EtMe$_3$Sb[Pd(dmit)$_2$]$_2$.

Finally, including our current work on EtMe$_3$Sb[Pd(dmit)$_2$]$_2$, no reproducible $\kappa_0/T$ was observed in any QSL candidates so far, despite that power-law temperature dependence of specific heat was observed in some cases. In this context, one may need to consider scenarios other than QSL for these frustrated magnetic systems. Recently, a random-singlet state is proposed based on the quenched disorder on spin-1/2 quantum magnets, which can induce the gapless QSL-like state \cite{RSjpsj,RSprx,RSnc}. The linear temperature dependence of specific heat may be an evidence for a power-law density of states of randomness \cite{RSprx,RSnc}. Interestingly, the linear or sublinear temperature dependence of our thermal conductivity $\kappa/T$ is in agreement with the theoretical prediction for a random-singlet state \cite{RSprx}. This power law is a consequence of the scattering of acoustic phonons by quantum two-level systems from the distribution of random singlets \cite{RSprx}, which may account for the huge suppression of our thermal conductivity shown in Fig. 4. However, we also notice that the specific heat is insensitive to the magnetic field and the susceptibility goes to a constant at low temperatures for EtMe$_3$Sb[Pd(dmit)$_2$]$_2$ \cite{dmitC,dmitprb08}, which are not consistent with the random singlet model.

In summary, we have revisited the thermodynamic and heat transport properties of triangular-lattice organic QSL candidate EtMe$_3$Sb[Pd(dmit)$_2$]$_2$. A linear term in the specific heat is well reproduced, as in previous report \cite{dmitC}, but the thermal conductivity shows a completely different behavior from Ref. \cite{dmitk}. No residual linear term $\kappa_0/T$ is observed at zero-temperature limit, suggesting the absence of mobile gapless fermionic excitations in EtMe$_3$Sb[Pd(dmit)$_2$]$_2$. A magnetic field of 6 T does not affect the thermal conductivity, and its absolute value is even 10 times smaller than the nonmagnetic reference compound Et$_2$Me$_2$Sb[Pd(dmit)$_2$]$_2$. We conclude that there is no magnetic thermal conductivity but only the phonon thermal conductivity in EtMe$_3$Sb[Pd(dmit)$_2$]$_2$, and the phonons are strongly scattered by the frustrated spins. The absence of reproducible $\kappa_0/T$ in any QSL candidates so far presents a direct challenge to the realization of a gapless QSL with highly mobile spinons in frustrated quantum magnets.

We are aware of a similar thermal conductivity study of EtMe$_3$Sb[Pd(dmit)$_2$]$_2$ single crystals by the Taillefer group in Sherbrooke \cite{Louis}. This work was supported by the Ministry of Science and Technology of China (Grant No: 2016YFA0300503 and 2015CB921401), the Natural Science Foundation of China (Grant No. 11421404), and the NSAF (Grant No: U1630248). It was also partially supported by the JSPS Grant-in-Aids for Scientific Research (S) (grant no. JP16H06346).\\

\noindent $^*$ E-mail: shiyan$\_$li$@$fudan.edu.cn


\begin{thebibliography}{99}

\bibitem{anderson73} P. W. Anderson, Resonating valence bonds: A new kind of insulator?,  Mater. Res. Bull. {\bf 8}, 153 (1973).
\bibitem{anderson87} P. W. Anderson, The resonating valence bond state in La$_2$CuO$_4$ and superconductivity, Science {\bf 235}, 1196 (1987).
\bibitem{balents10} L. Balents, Spin liquids in frustrated magnets, Nature (London) {\bf 464}, 199 (2010).
\bibitem{savary17} L. Savary and L. Balents, Quantum spin liquids: A review, Rep. Prog. Phys. {\bf 80}, 016502 (2017).
\bibitem{zhourmp} Y. Zhou, K. Kanoda, and T. K. Ng, Quantum spin liquid states, Rev. Mod. Phys. {\bf 89}, 025003 (2017).
\bibitem{ETprl03} Y. Shimizu, K. Miyagawa, K. Kanoda, M. Maesato, and G. Saito, Spin-Liquid State in an Organic Mott Insulator with a Triangular Lattice, Phys. Rev. Lett. {\bf 91}, 107001 (2003).
\bibitem{ETprb06} Y. Shimizu, K. Miyagawa, K. Kanoda, M. Maesato, and G. Saito, Emergence of inhomogeneous moments from spin liquid in the triangular-lattice Mott insulator $\kappa$-(BEDT-TTF)$_2$Cu$_2$(CN)$_3$, Phys. Rev. B {\bf 73}, 140407(R) (2006).
\bibitem{ETC} S. Yamashita, Y. Nakazawa, M. Oguni, Y. Oshima, H. Nojiri, Y. Shimizu, K. Miyagawa, and K. Kanoda, Thermodynamic properties of a spin-1/2 spin-liquid state in a $\kappa$-type organic salt, Nat. Phys. {\bf 4}, 459 (2008).
\bibitem{ETk} M. Yamashita, N. Nakata, Y. Kasahara, T. Sasaki, N. Yoneyama, N. Kobayashi, S. Fujimoto, T. Shibauchi, and Y. Matsuda, Thermal-transport measurements in a quantum spin-liquid state of the frustrated triangular magnet $\kappa$-(BEDT-TTF)$_2$Cu$_2$(CN)$_3$, Nat. Phys. {\bf 5}, 44 (2009).
\bibitem{dmitJPCM}T. Itou, A. Oyamada, S. Maegawa, M. Tamura, and R. Kato, Spin-liquid state in an organic spin-1/2 system on a triangular lattice, EtMe$_3$Sb[Pd(dmit)$_2$]$_2$, J. Phys.: Condens. Matter {\bf 19}, 145247 (2007).
\bibitem{dmitprb08} T. Itou, A. Oyamada, S. Maegawa, M. Tamura, and R. Kato, Quantum spin liquid in the spin-1/2 triangular antiferromagnet EtMe$_3$Sb[Pd(dmit)$_2$]$_2$, Phys. Rev. B {\bf 77}, 104413 (2008).
\bibitem{dmitk} M. Yamashita, N. Nakata, Y. Senshu, M. Nagata, H. M. Yamamoto, R. Kato, T. Shibauchi, and Y. Matsuda, Highly mobile gapless excitations in a two-dimensional candidate quantum spin liquid, Science {\bf 328}, 1246 (2010).
\bibitem{dmitNMR1} T. Itou, A. Oyamada, S. Maegawa, and R. Kato, Instability of a quantum spin liquid in an organic triangular-lattice antiferromagnet, Nat. Phys. {\bf 6}, 673 (2010).
\bibitem{dmitNMR2} T. Itou, K. Yamashita, M. Nishiyama, A. Oyamada, S. Maegawa, K. Kubo, and R. Kato, Nuclear magnetic resonance of the inequivalent carbon atoms in the organic spin-liquid material EtMe$_3$Sb[Pd(dmit)$_2$]$_2$, Phys. Rev. B {\bf 84}, 094405 (2011).
\bibitem{dmitC} S. Yamashita, T. Yamamoto, Y. Nakazawa, M. Tamura, and R. Kato, Gapless spin liquid of an organic triangular compound evidenced by thermodynamic measurements, Nat. Commun. {\bf 2}, 275 (2011).
\bibitem{dmittorque} D. Watanabe, M. Yamashita, S. Tonegawa, Y. oshima, H. M. Yamamoto, R. Kato, I. Sheikin, K. Behnia, T. Terashima, S. Uji, T. Shibauchi, and Y. Matsuda, Novel Pauli-paramagnetic quantum phase in a mott insulator, Nat. Commun. {\bf 3}, 1090 (2012).
\bibitem{YMGOsr} Y. S. Li, H. Liao, Z. Zhang, S. Li, F. Jin, L. Ling, L. Zhang, Y. Zou, L. Pi, Z. Yang, J. Wang, Z. Wu, and Q. Zhang, Gapless quantum spin-liquid ground state in the two-dimensional spin- 1/2 triangular antiferromagnet YbMgGaO$_4$, Sci. Rep. {\bf 5}, 16419 (2015).
\bibitem{YMGOneutron1} Y. Shen, Y. D. Li, H. Wo, Y. Li, S. Shen. B. Pan, Q. Wang, H. C. Walker, P. Steffens, M. Boehm, Y. Hao, D. L. Quintero-Castro, L. W. Harriger, M. D. Frontzek, L. Hao, S. Meng, Q. M. Zhang, G. Chen, and J. Zhao, Evidence for a spinon Fermi surface in a triangular lattice quantum-spin-liquid candidate, Nature (London) {\bf 540}, 559 (2016).
\bibitem{YMGOneutron2} J. A. M. Paddison, M. Daum, Z. Dun, G. Ehlers, Y. Liu, M. B. Stone, H. Zhou, and M. Mourigal, Continuous excitations of the triangular-lattice quantum spin liquid YbMgGaO$_4$, Nat. Phys. {\bf 13}, 117 (2017).
\bibitem{YMGOk} Y. Xu, J. Zhang, Y. S. Li, Y. J. Yu, X. C. Hong, Q. M. Zhang, and S. Y. Li, Absence of Magnetic Thermal Conductivity in the Quantum Spin-Liquid Candidate YbMgGaO$_4$, Phys. Rev. Lett. {\bf 117}, 267202 (2016).
\bibitem{YZGO} Z. Ma, J. Wang, Z. Y. Dong, J. Zhang, S. Li, S. H. Zheng, Y. Yu, W. Wang, L. Che, K. Ran, S. Bao, Z. Cai, P. Cermak, A. Schneidewind, S. Yano, J. S. Gardner, X. Lu, S. L. Yu, J. M. Liu, S. Y. Li, J. X. Li, and J. Wen, Spin-Glass Ground State in a Triangular-Lattice Compound YbZnGaO$_4$, Phys. Rev. Lett. {\bf 120}, 087201 (2018).
\bibitem{katoreview04} R. Kato, Conducting Metal Dithiolene Complexes: Structural and Electronic Properties, Chem. Rev. {\bf 104}, 5319 (2004).
\bibitem{katoreview11} K. Kanoda and R. Kato, Mott Physics in Organic Conductors with Triangular Lattices, Annu. Rev. Condens. Matter Phys. {\bf 2}, 167 (2011).
\bibitem{fit1} M. Sutherland, D. G. Hawthorn, R. W. Hill, F. Ronning, S. Wakimoto, H. Zhang, C. Proust, E. Boaknin, C. Lupien, L. Taillefer, R. X. Liang, D. A. Bonn, W. N. Hardy, R. Gagnon, N. E. Hussey, T. Kimura, M. Nohara, and H. Takagi, Thermal conductivity across the phase diagram of cuprates: Low-energy quasiparticles and doping dependence of superconducting gap, Phys. Rev. B {\bf 67}, 174520 (2003).
\bibitem{fit2} S. Y. Li, J. B. Bonnemaison, A. Payeur, P. Fournier, C. H. Wang, X. H. Chen, and L. Taillefer, Low-temperature phonon thermal conductivity of single crystalline Nd$_2$CuO$_4$: Effects of sample size and surface roughness, Phys. Rev. B {\bf77}, 134501 (2008).
\bibitem{RPP11} B. J. Powell and R. H. McKenzie, Quantum frustration in organic Mott insulators: from spin liquids to unconventional superconductors, Rep. Prog. Phys. {\bf 74}, 056501 (2011).
\bibitem{ktheory1} C. P. Nave and P. A. Lee, Transport properties of a spinon Fermi surface coupled to a U(1) gauge field, Phys. Rev. B {\bf 76}, 235124 (2007).
\bibitem{ktheory2} S. S. Lee and P. A. Lee, U(1) Gauge Theory of the Hubbard Model: Spin Liquid States and Possible Application to $\kappa$-(BEDT-TTF)$_2$Cu$_2$(CN)$_3$, Phys. Rev. Lett. {\bf 95}, 036403 (2005).
\bibitem{ktheory4} Yi Zhou and Tai-Kai Ng, Spin liquid states in the vicinity of a metal-insulator transition, Phys. Rev. B {\bf 88}, 165130 (2013).
\bibitem{ktheory3} Y. Werman, S. Chatterjee, S. C. Morampudi, and E. Berg, Signatures of fractionalization in spin liquids from interlayer transport, Phys. Rev. X {\bf 8}, 031064 (2018).
\bibitem{kreview} M. Yamashita, T. Shibauchi, and Y. Matsuda, Thermal-transport studies on two-dimensional quantum spin liquids, Chem. Phys. Chem. {\bf 13}, 74 (2012).
\bibitem{RSjpsj} K. Watanabe, H. Kawamura, H. Nakano, and Toru Sakai, Quantum Spin-Liquid Behavior in the Spin-1/2 Random Heisenberg Antiferromagnet on the Triangular Lattice, J. Phys. Soc. Jpn. {\bf 83}, 034714 (2014).
\bibitem{RSprx} I. Kimchi, A. Nahum, and T. Senthil, Valence Bonds in Random Quantum Magnets: Theory and Application to YbMgGaO$_4$, Phys. Rev. X {\bf 8}, 031028 (2018).
\bibitem{RSnc} I. Kimchi, J. P. Sheckelton, T. M. McQueen, and P. A. Lee, Scaling and data collapse from local moments in frustrated disordered quantum spin systems, Nat. Commun. {\bf 9}, 4367 (2018).
\bibitem{Louis} P. Bourgeois-Hope, F. Laliberte, E. Lefrancois, G. Grissonnanche, S. Rene de Cotret, R. Gordon, R. Kato, L. Taillefer, and N. Doiron-Leyraud, arXiv:1904.10402.

\end{thebibliography}
\end{document}